\documentclass[prl,twocolumn,showpacs]{revtex4}
\usepackage{amsfonts,amsmath,mathrsfs,epsfig,amsbsy,bm,verbatim,color}

\newcommand {\beq} {\begin{equation}}
\newcommand {\eeq} {\end{equation}}
\newcommand {\bqa} {\begin{eqnarray}}
\newcommand {\eqa} {\end{eqnarray}}

\begin{document}
\title{Chiral Rashba spin textures in ultra-cold Fermi gases}
\author{Jay D. Sau$^1$}
\author{Rajdeep Sensarma$^{1}$}
\author{Stephen Powell$^{1}$}
\author{I. B. Spielman$^{1,2}$}
\author{S. Das Sarma$^1$}

\affiliation{$^1$Condensed Matter Theory Center and Joint Quantum Institute, Department of Physics, University of
Maryland, College Park, Maryland 20742-4111, USA\\
$^2$National Institute of Standards and Technology, Gaithersburg,
 Maryland, 20899.}
\begin{abstract}
Spin-orbit coupling is an important ingredient in many recently discovered
 phenomena such as the spin-Hall effect and topological
 insulators. Of particular interest is topological superconductivity, with its
 potential application in topological quantum computation. The 
absence of disorder in ultra-cold atomic systems makes them ideal 
 for quantum computation applications, however, the 
spin-orbit (SO) coupling schemes proposed thus far are experimentally impractical owing to large 
spontaneous emission rates in the alkali fermions.
 In this paper, we develop a scheme to
 generate Rashba SO coupling with a low spontaneous emission extension to 
  a recent experiment.
 We show that this scheme generates a Fermi surface spin texture 
for $^{40}\rm{K}$ atoms, which is observable in 
time-of-flight measurements. The chiral spin texture, together with conventional 
$s$-wave interactions leads to
 topological
 superconductivity and non-Abelian Majorana quasiparticles.
\end{abstract}
\maketitle
\paragraph{Introduction:}
The physics of coupled  spin and motional degrees of freedom, and 
associated phenomena such as the spin-Hall effect \cite{murakami,sinova}, 
are a subject of intense interest: first in semiconductors \cite{orenstein} and now
 in cold-atom systems \cite{zoller,duan,tripod}.
  Recent predictions   
 of new topological states of matter based on spin-orbit (SO) 
coupling such as topological insulators \cite{kane} and topological superconductors \cite{schnyder} 
have generated great excitement  both  experimentally and theoretically.
In solids, SO coupling occurs naturally in systems with broken inversion symmetry, while SO coupling must
 be engineered for  cold atoms,  where laser fields  break inversion symmetry. 

Time-reversal(TR) symmetry breaking
 topological superconductor \cite{schnyder,sau,longprb} are one 
 particularly interesting class of systems that may be realized in the presence of 
 SO coupling.
 These support Majorana quasiparticle excitations which obey non-Abelian statistics \cite{Wilczek2}.
 Systems exhibiting  non-Abelian statistics have topologically degenerate ground states
 that can potentially be used for topological quantum computation \cite{Kitaev,dassarma}.
 The promise of realizing the elusive Majorana particles  
together with the topological ground state degeneracy in superconducting 
systems \cite{Read} has inspired considerable experimental effort studying superconductivity in SO coupled solid state systems in magnetic
 fields \cite{franz,Wilczek-3}.

\begin{figure}[tbp]
\begin{center}
\includegraphics[width=0.45\textwidth,angle=0]{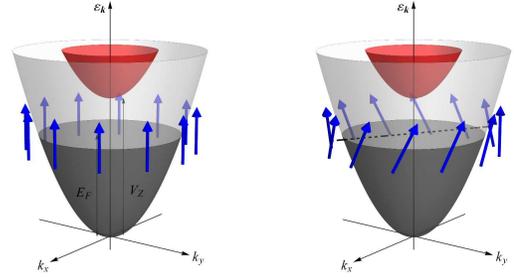}
\end{center}
\caption{(a)  Uniform Fermi surface spin texture in a system of fermions with Zeeman splitting $V_Z$.
 The Fermi surface is taken to be in the lower band ( spin polarized-up)
 with  Fermi energy $E_F$.
 (b) Chiral Fermi surface spin texture in a system of fermions with Zeeman coupling
 $V_Z$ and Rashba SO coupling.
 The Rashba SO coupling causes a counter-clockwise 
canting of spins (canted arrows)  on the otherwise 
spin-up Fermi surface. The canting-induced overlap 
between time-reversed spins at opposite points $\vec k$ and $-\vec k$ 
on the Fermi surface allows fermion pairing by a microscopic   
$s$-wave interaction leading to $p\!-\!$wave topological superfluidity.} 
\label{Fig0}
\end{figure}

 To understand the role of SO coupling in topological 
superfluidity \cite{chuanwei}, consider a  system of  
fermionic atoms in a Zeeman field $V_Z> E_F,k_B T$,  with  attractive $s$-wave interactions [Fig.~\ref{Fig0}(a)].
Pairing is in the
 spin-singlet channel, i.e., with opposite spins.
  Here $T$  is the temperature, and $E_F=\hbar^2 k_F^2/2M$
 is the Fermi energy of atoms with mass $M$ and Fermi 
momentum $\hbar k_F$. Since $V_Z>E_F$, 
 the fermions are  completely spin polarized  and the resulting state is an ideal  non-interacting
 Fermi gas. 
 On the other hand, in the presence of Rashba SO coupling [$H_R=\alpha(\bm\sigma_x p_y-\bm\sigma_y p_x)$] with strength
 $\alpha$ ,
 where $\bm\sigma_{x,y,z}$ are Pauli matrices,
 the spins at the Fermi-surface are canted
 [Fig. 1(b)]. Since pairs of fermions at momenta 
 $\hbar\vec k$ and $-\hbar\vec k$ on the Fermi surface have a component
 in the singlet-channel \cite{chuanwei,longprb}, 
 an attractive $s$-wave interaction with coupling constant
 $g$ generates an effective $p$-wave interaction $\tilde{g}\sim g\alpha k_F/V_Z$ at the Fermi surface 
.
 This interaction leads to 
pairing and a superfluid quasiparticle gap. 
Such a completely gapped superfluid state occupying a single Fermi surface 
has been shown to be topological with zero-energy Majorana modes 
as low-energy excitations of the system \cite{sau}. Thus SO coupling 
together with Zeeman splitting would lead to topological superconductivity 
with Majorana fermions. Unlike topological phases such as 
topological insulators, toplogical superfluids are created by arbitrarily weak SO coupling \cite{sau}. 
 The strength of SO coupling  determines 
the size of the gap and thus the 
robustness of the topological phase.

The three ingredients needed to realize TR breaking
 topological superconductors are: ordinary $s$-wave interactions, 
Zeeman splitting, and SO coupling \cite{sau,longprb}.
 The first two ingredients are already well-established in 
cold atomic systems. Several schemes  for realizing SO coupling in
 cold atomic systems \cite{zoller,tripod,duan}, where state-dependent laser fields lead to
  effective SO-coupled Hamiltonians for the laser-dressed atoms, 
 are the basis of proposals for topological systems such as topological insulators 
\cite{tudor} and topological superconductors \cite {chuanwei, fujimoto}.
 Despite the theoretical interest in such  SO coupling schemes, 
they have thus far eluded experiment. One reason
for this is that the simpler schemes such as the 
tripod-scheme \cite{tripod} require  resonant 
coupling to achieve the required SO coupling, and spontaneous 
emission from nearby excited states in actual experimental systems becomes prohibitive.

An exciting recent development  is the first experimental 
realization of a restricted class of SO coupled 
Hamiltonians  in a bosonic cold  atomic system \cite{spielman,spielman1}.
This experiment used the three levels in the $^{87}\rm{Rb}$ $F=1$ 
ground state manifold, and reduces spontaneous emission 
by working in the far-detuned limit \cite{spielman1,stamperkurn}.
 In addition, by using the lowest energy manifold of dressed
 states collisional decay was eliminated \cite{spielmanpra}.
The resulting SO coupling, \cite{spielman2} 
 $H_{R+D}\propto \bm\sigma_y p_x$ can in principle 
lead to a superfluid gap. Unfortunately, for fermions this 
 equal sum of Rashba and Dresselhaus SO coupling results in
a non-topological gapless $p_x$-superfluid. 

Here, we design a cold-atom system with Rashba SO coupling 
using an extension of the existing setup used to realize $H_{R+D}$.
 The SO coupling here is a function of the intensity of the applied lasers, while, in previous schemes, the SO coupling 
is geometric and  independent of laser strength for large 
enough coupling \cite{tripod}.
Although all Raman coupling schemes are 
susceptible to spontaneous emission, our scheme uses small 
two-photon coupling, minimizing spontaneous emission. Atom-chip based systems, where 
spatially varying magnetic fields replace Raman couplings,
 have recently been proposed to realize systems such as topological
 insulators that require strong SO coupling \cite{spielman2}.

Following the description of our proposed experimental setup, we verify analytically 
that  the effective Hamiltonian in the weak  Raman coupling limit has the requisite Rashba SO 
coupling form. Going beyond this limit, we show by  direct numerical
 calculation that the eigenstates have the chiral momentum-space spin texture required to
 create $p$-wave interactions with no nodes.
 Varying the relative laser intensities allows us to 
tune the SO coupling between the $H_{R}$ (Rashba) and $H_{R+D}$ forms \cite{orenstein}.
 The spin texture calculation predicts the results of time-of-flight measurements of
 spin-density, which would directly establish that the 
necessary effective Hamiltonian to create the Majorana particle 
has been created in the atomic Fermi gas (all one needs to do is to add the 
$s$-wave superfluidity through the appropriate Feshbach resonance.).

\begin{figure}[tbp]
\begin{center}
\includegraphics[width=0.4\textwidth,angle=0]{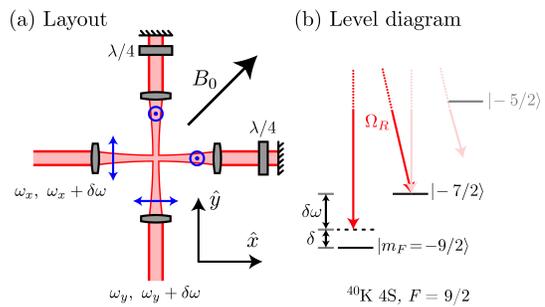}
\end{center}
\caption{(a) Laser fields applied to an ultra-cold $^{40}K$ gas
 generate a Rashba SO coupling between the 
$m_F=-7/2$ and $m_F=-9/2$ states. Effective Zeeman couplings 
that vary along $\hat x$ and $\hat y$ directions are generated by 
Raman-coupling these states through an  
 excited state $e$ by lasers of frequencies $\omega_x,\omega_x+\delta \omega$ along $\hat x$ and
  $\omega_y,\omega_y+\delta \omega$ along $\hat y$.
Appropriately tuning the phases of these lasers at the atomic 
gas generates Rashba and Dresselhaus SO coupling.
(b) Atomic level structure (solid lines) of an effective spin-1/2 atom with
 hyperfine states $|\sigma_z=+1\rangle\equiv |m_F=-7/2\rangle$ and 
$|\sigma_z=-1\rangle\equiv |m_F=-9/2\rangle$.
 The detuning $\delta$ from the state $|\sigma_z=+1\rangle$ sets the position 
independent Zeeman coupling along $\bm\sigma_z$.} 
\label{Fig1}
\end{figure}

\paragraph{Setup:}  We propose to illuminate a sample of quantum
 degenerate fermionic $^{40}\textrm {K}$ in its electronic ground state
 with two pairs of counterpropagating ``Raman'' lasers.  These beams are
 far-detuned about $\Delta\sim 2$ THz to the red of the $4\textrm{S}_{1/2}$ to $4\textrm{P}_{1/2}$ (D1)
 transition $(\sim 400 \textrm{ THz})$, and each beam is composed of two frequencies
 $\omega_{x,y}$ and $\omega_{x,y}+\delta\omega$ which drive transitions between
 different sub-levels of the ground state $F=9/2$ hyperfine manifold.  As pictured in Fig.~\ref{Fig1},
 a magnetic field $B_0\approx200$ G along $\hat x+\hat y$ resolves the different
 Zeeman sub-levels, and owing to a sizable quadratic Zeeman shift only one pair
 of states is Raman-resonant, here 
 $|m_F=-9/2\rangle$ and $|m_F=-7/2\rangle$. We will refer to these 
states as pseudo-spin $|\sigma_z=-1\rangle$ and $|\sigma_z=+1\rangle$ respectively. 
The excited states $4\textrm{P}_{1/2}$ and $4\textrm{P}_{3/2}$
 through which the 
Raman coupling between $|\sigma_z=\pm 1\rangle$ is driven will be 
collectively referred to as $e$. 

\paragraph{Effective Zeeman coupling:}
The two pairs of lasers in our setup along 
 $\hat x$ and $\hat y$ respectively, are detuned from the excited 
state $e$ by slightly different detunings
 $\Delta_x$ and $\Delta_y$, so that 
they induce independent Raman couplings.
The Rabi coupling between states $|\sigma_z=\pm 1\rangle$ 
 and the excited state $e$,  driven  
by the pair of lasers along  $a=x\textrm{ or }y$ together with their 
reflected counterparts,
 can be written as a complex two-component 
spinor amplitude $\bm f^{(a)}_{\sigma_z}(\vec r)$.
 The spinor potential  $\bm f^{(a)}(\vec r)= \Gamma^{(a)}
[\bm f^{(a,+)} e^{i \vec G^{(a)}\cdot\vec r}+\bm f^{(a,-)} e^{-i \vec G^{(a)}\cdot\vec r}]$ results from lasers traveling in opposite 
directions; the spinors $\bm f^{(a,\pm)}$ 
encode the relative phases of the lasers along $\pm \hat{a}$ 
and at frequencies $\omega_a$ and $\omega_a+\delta\omega$.
 Here $\Gamma^{(a)}$ is the overall coupling 
constant along $\hat{a}$.

In the far-detuned limit, where all energy eigenvalues
 of states of interest are
 much smaller than the detuning $\Delta_a$,
 each pair of lasers 
generates an effective spatially dependent 
Zeeman potential within the rotating wave approximation (RWA),  given by 
\begin{align}
&\bm F^{(a)}(\vec r)=\bm f^{(a)}(\vec r)\bm f^{(a)\dagger}(\vec r)/\Delta_a.
\end{align}
To generate Rashba and Dresselhaus SO coupling 
we choose $\bm f^{(x,\pm)}$ to be eigenvectors  
eigenvectors of $\bm\sigma_y$ (i.e., $\bm\sigma_y \bm f^{(x,\pm)}=\pm \bm f^{(x,\pm)}$) and  $\bm f^{(y,\pm)}$ to be 
eigenvectors of $\bm\sigma_x$ (i.e., $\bm\sigma_x \bm f^{(y,\pm)}=\mp \bm f^{(y,\pm)}$).
 With this choice the effective Zeeman potential $\bm F(\vec r)=\sum_a \bm F^{(a)}(\vec r)$ is 
\begin{align}
&\bm F(\vec r)=\Omega_R^{(y)}\sin{\left(G_0 y\right)}\bm\sigma_y-\Omega_R^{(x)}\sin{\left(G_0 x\right)}\bm\sigma_x\nonumber\\
&+\left[\Omega_R^{(x)}\cos{(G_0 x)}+\Omega_R^{(y)}-\cos{(G_0 y)}\right]\bm\sigma_z  \label{eq:Zeeman}
\end{align}
where $\Omega_R^{(a)}=\Gamma^{(a)2}/\Delta^{(a)}$ is the effective Raman 
coupling in the direction $a$. 
 Here we have taken $\vec G^{(x)}=G_0\hat x/2$ and $\vec G^{(y)}=G_0\hat y/2$
where $G_{0}=(2\omega_a+\delta\omega)/c=4\pi/\lambda$ and 
$\lambda\sim 0.7\quad\mu$m is the wavelength of light
  \footnote{We have neglected 
the slightly different magnitudes of the wavevectors $\vec G^{(a)}_{\beta}$ corresponding to 
$\omega_a$ and $\omega_a+\delta\omega$.}.
In Eqn.~\ref{eq:Zeeman}, we ignored a constant and irrelevant spin-independent shift.
From the above equations, we find that only the relative phases of the beams propagating in the same direction 
(with the same $a$) contribute to the effective Zeeman coupling. Furthermore, the only effect of 
changing the relative phase of counterpropagating lasers (e.g $+\hat x$ and $-\hat x$)
is to shift the origin of the system.
\paragraph{Bloch Hamiltonian:}
 Since the effective position dependent Zeeman potential 
$\bm F(\vec r)$ is periodic in space, the Hamiltonian
 for our system 
 has a discrete translational symmetry.
 Using Bloch's theorem, the spinor eigenstate of the Hamiltonian
 of the system can be expanded as 
$\bm \phi^{(n)}_{\vec k}(\vec r)= \sum_{\vec G}\bm C^{(n)}_{\vec k+\vec G} e^{\imath(\vec k+\vec G)\cdot \vec r}$ such that the eigenstate
 $\bm\phi^{(n)}_{\vec k}$
 has a conserved wavevector $\vec k$ defined modulo  
 reciprocal lattice vectors $\vec G=n_1\vec G^{(x)}+n_2\vec G^{(y)}$.
The Bloch eigenvectors $\bm \phi^{(n)}_{\vec k}$ at wavevector
 $\vec k$ are eigenvectors of the Bloch Hamiltonian at 
momentum $\hbar\vec k$ 
\begin{equation}
\bm H_{\vec{k}}(\vec k+\vec G;\vec k+\vec G')=\left(\frac{\hbar^2|\vec k+\vec G|^2}{2 M}\bm 1+\frac{\delta}{2}\bm\sigma_z\right)\delta_{\vec G,\vec G'}+\bm F(\vec G-\vec G')\label{eq:Hbloch}
\end{equation}
where $\bm F(\vec G)$ is the Fourier transform of the laser-induced 
spatially varying Zeeman coupling $\bm F(\vec r)$.  Here $\delta$ is the detuning induced Zeeman splitting    
 obtained by choosing the detuning of the respective lasers
as shown in Fig.~\ref{Fig1}(b).
 For $\vec k$ restricted to the 
first Brillouin zone (FBZ), 
the set of Bloch eigenvectors $\bm \phi^{(n)}_{\vec k}$, 
 labeled by band-index $n$, and corresponding energy eigenvalues
 $\epsilon^{(n)}_{\vec k}$ 
can be determined using 
 $\bm H_{\vec k}\bm\phi^{(n)}_{\vec k}=\epsilon^{(n)}_{\vec k}\bm\phi^{(n)}_{\vec k}$.

\paragraph{Small $\Omega_R$ limit:}
The explicit Rashba-Dresselhaus form of the effective Hamiltonian for the two lowest bands (which serve 
as a pseudo-spin basis) 
 can be derived in the limit of large recoil energy
 $E_R=\frac{h^2}{2 M \lambda^2}\approx h \times 8.4 \textrm{ kHz}\gg \delta,\Omega_R$ 
for $\lambda\approx 770$ nm. 
 Absent Raman coupling $(\Omega_R=0)$, the Hamiltonian $\bm H_{\vec k}$ is diagonal in the spin
 and reciprocal lattice-vector $(\vec G)$ basis.
 In this case, for wavevectors $\vec k$ in the FBZ, the lowest pair of bands  contain
 only states with  $\vec G=0$, while the higher bands, separated in energy by at least $E_R$,
 contain states with $\vec G\neq 0$. 

A small Raman coupling $\Omega_R\ll E_R$  couples the unperturbed $(\Omega_R=0)$ system 
such that the Schr\"odinger equation is written  as:   
\begin{align}
&[ k^2\bm 1+\frac{\delta}{2}\bm\sigma_z-\epsilon^{(n)}_{\vec k}\bm 1]\bm C^{(n)}_{\vec k}+\sum_{\vec G\neq 0} \bm F(-\vec G)\bm C^{(n)}_{\vec k+ \vec {G} }=0\label{eq:SE1}\\
&\left[|\vec k+\beta \vec G_\gamma|^2-\epsilon^{(n)}_{\vec k}\right]\bm C^{(n)}_{\vec k+\vec G}+\bm F(\vec G) \bm C^{(n)}_{\vec k}=0\label{eq:SE2},
\end{align}
where we have set $\hbar^2/2 M=1$, and 
 have ignored the contribution of $\vec G\neq 0$ to the second term 
in Eq.~\ref{eq:SE2} which is of order $\left(\Omega_R/E_R\right)^3$.
 Furthermore, we ignored the contribution 
$\bm F(\vec G=0)$ to the equation, 
which leads to an overall energy shift. The unperturbed  states in the higher bands,  
with a  small contribution, can be 
adiabatically eliminated by 
substituting $\bm C^{(n)}_{\vec k+\vec G}$ from Eq.~\ref{eq:SE2}
 into Eq.~\ref{eq:SE1}; we obtain the effective
 Schr\"odinger equation 
\begin{equation}
\left[(k^2-\epsilon^{(n)}_{\vec k})\bm 1+
\frac{\delta}{2}\bm\sigma_z- \sum_{\vec G\neq 0}
\frac{\bm F(-\vec G)\bm F(\vec G)}{|\vec k+\vec G|^2-\epsilon^{(n)}_{\vec k}}\right]\bm\phi_{\vec k}=0.
\end{equation}
For $\epsilon^{(n)}_{\vec k}\ll E_R$, the $\epsilon^{(n)}_{\vec k}$ dependence of the last 
term is negligible.

 The effective Schr\"odinger equation (Eq.~\ref{eq:SE2}) for the lowest pair of bands in the FBZ takes the familiar Rashba form 
\begin{align}
&\left[(k^2-\mu-\epsilon^{(n)}_{\vec k})\bm 1+\frac{\delta}{2}\bm\sigma_z+\frac{\Omega_R^2}{2 E_R G_0}(k_y\bm\sigma_x-k_x\bm\sigma_y)\right]
\bm\phi^{(n)}_{\vec k}=0\label{eqrashbafinal},
\end{align}
when $\Omega^{(x)}_R=\Omega^{(y)}_R=\Omega_R$.
Pure Dresselhaus coupling is obtained by interchanging the spinors $\bm f^{(x,\pm)}$, 
while varying the ratio of $\Omega^{(x)}_R/\Omega^{(y)}_R$
  tunes  SO coupling from the  pure Rashba 
type to the $R+D$  type.

\begin{figure}[tbp]
\begin{center}
\includegraphics[width=0.5\textwidth]{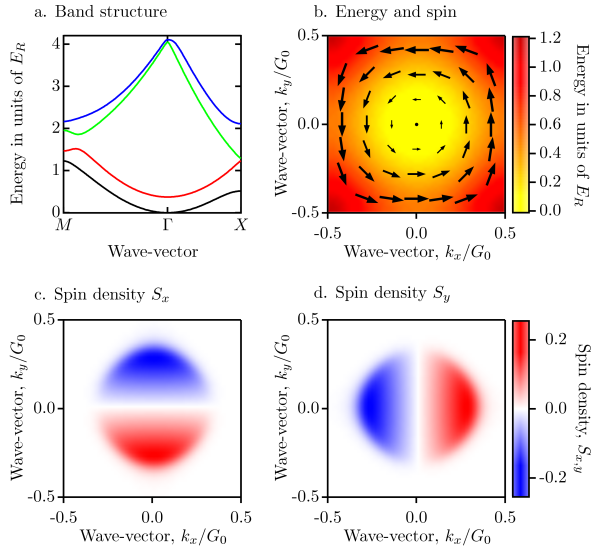}
\end{center}
\caption{(a) Band structure for the proposed system for the parameters 
 $\Omega_R=\Gamma^2/\Delta=0.8 E_R$ together with $\delta=0.4 E_R$. 
For $^{40}$K, the recoil energy  $E_R= h\times 8.4$ kHz.
 The points $\vec k=(k_x,k_y)=(0.5,0.5)G_0,(0,0)$
 and $(0.5,0)G_0$ have been labelled $M$, $\Gamma$ and $X$ on the horizontal-axis 
respectively.
 For these parameters, the lowest energy-band is  non-degenerate
 and isotropic.
(b) Dispersion and spin texture in the lowest band, showing nearly circular
constant-energy contours for small wavevectors (color scale) and chiral
spin texture characteristic of Rashba SO coupling (arrows indicate $x$-$y$ spin components).
 (c) Momentum  $S_x(\vec k)$ spin density. 
The corresponding time-of-flight phase-contrast measurement of $S_x$ spin density 
should show a $p_y$ symmetry. 
 (d)  Momentum $S_y(\vec k)$ spin density. 
The corresponding time-of-flight phase-contrast 
measurement of $S_y$ spin density 
should show a $p_x$ symmetry.} 
\label{Fig2}
\end{figure}

\paragraph{Numerical solution:}
We verified the Rashba coupling obtained from the perturbative calculation
 by numerically 
diagonalizing the Hamiltonian  (Eq.~\ref{eq:Hbloch}). 
 As is clear from the representative band structure shown in Fig.~\ref{Fig2}(a),
 there is a non-degenerate band at the lowest energy.
For these parameters, a reasonably high density Fermi gas  $(n\sim 1.2 \lambda^{-2})$,
at a temperature of $k_B T=E_F/8\approx 0.04 E_R$, would be in the single Fermi surface 
limit with  a Fermi energy of $E_F\approx 0.32 E_R$. 
.
The spin texture  in the occupied band, shown in Fig.~\ref{Fig2}(b-d), 
 is calculated from the spin-expectation value  
\begin{equation}
\vec{s}(\vec k)=\sum_{\vec G}\bm C^\dagger_{0 \vec k+\vec G}\vec{\bm\sigma} \bm C_{0\vec k+\vec G}
\end{equation} 
and is chiral, confirming the Rashba SO coupling.

\paragraph{Time-of-flight phase-contrast detection:}
 Our calculations [Fig. 3(c,d)] show a 
momentum-dependent chiral spin texture such that the $S_x$  and 
 $S_y$ spin densities have  $p_y$ and $p_x$ symmetries in momentum space 
respectively.
The chiral spin texture at the Fermi surface can be detected using 
phase-contrast imaging \cite{stamperkurn-1} following 
time-of-flight (TOF) expansion, where the final position (reflecting 
the initial momentum) will be correlated with initial spin.
 Phase-contrast imaging of the atoms can measure spin densities along 
arbitrary directions allowing one to measure $S_x$ and $S_y$ directly.

\paragraph{Conclusion:}
We proposed a precise, and experimentally feasible, scheme to generate Rashba SO coupling
 which eliminates the  heating problem of the tripod scheme due to spontaneous emission. 
We show by direct calculations that our proposed scheme should lead to the observation of chiral 
spin textures using a phase-contrast technique.
 The spin texture together with conventional 
$s$-wave inter-atomic interactions should lead to effective $p$-wave pairing
 and hence topological superconductivity and non-Abelian Majorana fermions.

This work was supported by DARPA-QuEST, JQI-NSF-PFC, LPS-NSA and 
ARO's atomtronics MURI.

\end{document}